\documentclass[aps,prl,reprint,superscriptaddress,showpacs,amsmath,amssymb,preprintnumbers]{revtex4-1}

\usepackage{graphicx}
\usepackage{dcolumn}
\usepackage{bm}
\usepackage[dvips]{color}

\begin{document}

\preprint{Revision 2}

\title{Origin and tailoring of the antiferromagnetic domain structure in $\alpha$-Fe$_2$O$_3$ thin films unraveled by statistical analysis of dichroic spectro-microscopy (X-PEEM) images}


\author{Odile Bezencenet}
\altaffiliation{THALES Research \& Technology France, 1, avenue Augustin Fresnel, F-91767 Palaiseau cedex, FRANCE}
\affiliation{CEA, IRAMIS, SPCSI, F-91191 Gif sur Yvette, France}
\author{Daniel Bonamy}
\affiliation{CEA, IRAMIS, SPCSI, F-91191 Gif sur Yvette, France}
\author{Rachid Belkhou}
\affiliation{Synchrotron SOLEIL, L'Orme des Merisiers Saint-Aubin, BP 48, F-91192 Gif-sur-Yvette Cedex, France}
\author{Philippe Ohresser}
\affiliation{Synchrotron SOLEIL, L'Orme des Merisiers Saint-Aubin, BP 48, F-91192 Gif-sur-Yvette Cedex, France}
\author{Antoine Barbier}
 \email[]{Contact author: abarbier@cea.fr}
\affiliation{CEA, IRAMIS, SPCSI, F-91191 Gif sur Yvette, France}


\date{\today}

\begin{abstract}
 The magnetic microstructure and domain wall distribution of antiferromagnetic $\alpha$-Fe$_2$O$_3$ epitaxial layers is determined by statistical image analyses. Using dichroic spectro-microscopy images, we demonstrate that the domain structure is statistically invariant with thickness and that the antiferromagnetic domain structure of the thin films is inherited from the ferrimagnetic precursor layer one, even after complete transformation into antiferromagnetic $\alpha$-Fe$_2$O$_3$. We show that modifying the magnetic domain structure of the precursor layer is a genuine way to tune the magnetic domain structure and domain walls of the antiferromagnetic layers.
\end{abstract}


\maketitle


Modern spintronics devices \cite{Chappert:NatureMaterials2007} are based on the use of a limited number of magnetic properties including ferromagnetic/antiferromagnetic (FM/AF) interfacial exchange bias  \cite{Magnan:Phys.Rev.Lett.2010,Nolting:Nature2000,Scholl:Science2000,Meiklejohn:Phys.Rev.1956}, spin torque currents \cite{Thomas:Nature2006} and domain wall motion \cite{Urazhdin:Phys.Rev.Let.2007,Thomas:Science2010} used \emph{e.g.} in race track memory cells \cite{Parkin:Science2008}. Antiferromagnets are expected to play an active role in this generation of devices but the investigation, observation and tuning of their domains remains unchallenged because they do not carry any net external magnetic dipole moment. The nanoscale control of magnetic \cite{Takamura:NanoLett.} and/or electric \cite{Chu:NanoLett.2009,Catalan:Phys.Rev.Lett.2008} domains in devices is of tremendous current interest because of the expected high industrial impact. Additionnaly, the magnetic domain microstructure in magnetic exchange coupled devices determines the level of electronic (Barkhausen) noise \cite{Barkhausen:Phys.Z.1919}, originating from magnetic domain wall motion, which is detrimental to most magneto-resistive sensors \cite{Shpyrko:Nature2007}. The investigation and observation of AF domains \cite{Scholl:Science2000,Hellwig:NatureMaterials2003,Sato:Nature2004,Kuch:NatureMaterials2006,Bode:NatureMaterials2006} has only been made possible recently by photoelectrons microscopies. Nevertheless, the detailed description and manipulation of the microstructure of AF domains, domain shape and size distribution still remain in infancy. More generally, extracting global information from sets of samples with different histories requires the development of specific tools. Here we demonstrate, using statistical physics methods applied to spectromicroscopy images \cite{Bauer:Rep.Prog.Phys.1994}, that the magnetic domain structure of an AF thin film is inherited from the magnetic domain structure of the ferri-magnetic precursor layer that exists at small layer thicknesses, before the AF spin ordering sets in, and remains unchanged afterwards, even after complete FM$\rightarrow$AF transformation. Modifying the magnetic domain structure of the ferrimagnetic precursor layer appears as a new and promising route to tailor the magnetic domain structure of AF layers having a ferro- or ferri-magnetic parent phase.

Our samples were hematite ($\alpha$-Fe$_2$O$_3$) layers in the 2-30 nm thickness range, that is below the probing depth of our analysis, grown on Pt(111) substrates. They were prepared by atomic-oxygen-plasma-assisted molecular beam epitaxy with the substrate held at $\sim$750 K and a deposition rate of 2 {\AA}/min \cite{Barbier:Phys.Rev.B2005}. They have a number of crucial advantages: (i) hematite layers can easily be cleaned after air exposure; (ii) Pt(111) single crystals promote high quality epitaxial thin film growth \cite{Barbier:Mat.Sci.andEng.B2007} and (iii) metallic substrates provide an efficient way to overcome charge build-up effects induced otherwise by insulating hematite layers which hinder the use of electron based techniques. Previous X-ray Magnetic Circular and Linear Dichroism (XMCD and XMLD) measurements revealed a  ferrimagnetic $\gamma$-Fe$_2$O$_3$(111) layer with spinel structure up to a thickness of 3 nm whereas pure  AF $\alpha$-Fe$_2$O$_3$(0001), with corundum structure, is obtained above \cite{Barbier:Phys.Rev.B2005}. Both phases were not found to coexist for any sample. With respect to these properties our $\alpha$-Fe$_2$O$_3$ films can be considered prototypical for corundum structured magnetic metal oxides.

High-resolution magnetic imaging experiments were performed on the Nanospectroscopy beamline at ELETTRA synchrotron (Trieste, Italy) and the SIM beamline at SLS synchrotron (Villigen, Switzerland), using an Elmitec GmBH commercial LEEM/PEEM microscope \cite{Bauer:Rep.Prog.Phys.1994}. The AF micromagnetic spin structure of the hematite surface has been determined taking advantage of the large XMLD effect associated with the Fe $L_{2,3}$ edges using vertical/horizontal polarized light \cite{Nolting:Nature2000}. In the XMLD-PEEM method, the electron yield difference between the two main multiplet lines (called, by convention, L$_{2A}$ and L$_{2B}$ with respect to increasing photon energies) is proportional to the scalar product of the AF axis and the X-ray polarization vector \cite{Brice-Profeta:J.Magn.Magn.Mater.2005}. When the linear polarization lies in the surface plane of the sample, it enables to essentially map in-plane component of AF axis. The ferrimagnetic domain structure of $\gamma$-Fe$_2$O$_3$ was obtained using circular dichroic images derived from incident right- and left- handed circularly polarized photons, exploiting the XMCD  effect at the Fe L$_3$ edge \cite{Nolting:Nature2000,Scholl:Science2000}.

\begin{figure}
 \includegraphics[angle=0, width=8.5cm]{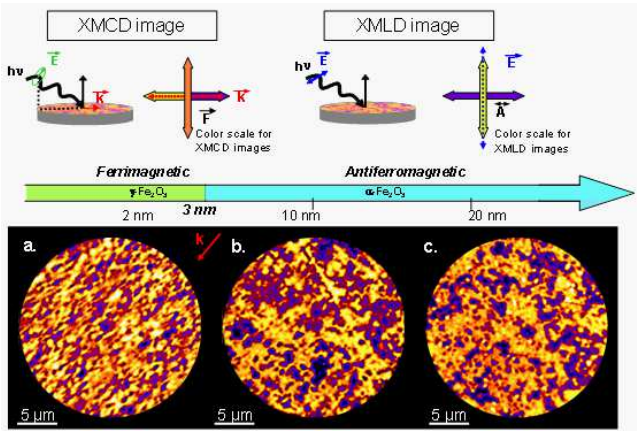}%
\caption{\label{general}(Color online) 20 $\mu$m field of view X-PEEM images recorded at the Fe L$_{2A,B}$-edge for different Fe$_2$O$_3$ film thicknesses: (a) 2 nm $\gamma$-Fe$_2$O$_3$; (b) 10 nm $\alpha$-Fe$_2$O$_3$; (c) 20 nm $\alpha$-Fe$_2$O$_3$. The incident x-ray beam was oriented at 74° with respect to the surface normal direction. The contrast mainly arises from ferrimagnetic domains in  $\gamma$-Fe$_2$O$_3$ and from AF domains in  $\alpha$-Fe$_2$O$_3$ using respectively XMCD and XMLD dichroic images at the Fe L$_3$ edge and the multiplet structure at the Fe L$_2$ edge respectively. The upper panel of the figure illustrates the relationship between the contrast scale of the images and the in-plane projection of the ferrimagnetic axis \textbf{F} (respectively AF axis \textbf{A}) with the X-ray propagation direction \textbf{k} (respectively the photon polarization \textbf{E}). }
\end{figure}

\begin{figure}
\includegraphics[width=0.75\columnwidth]{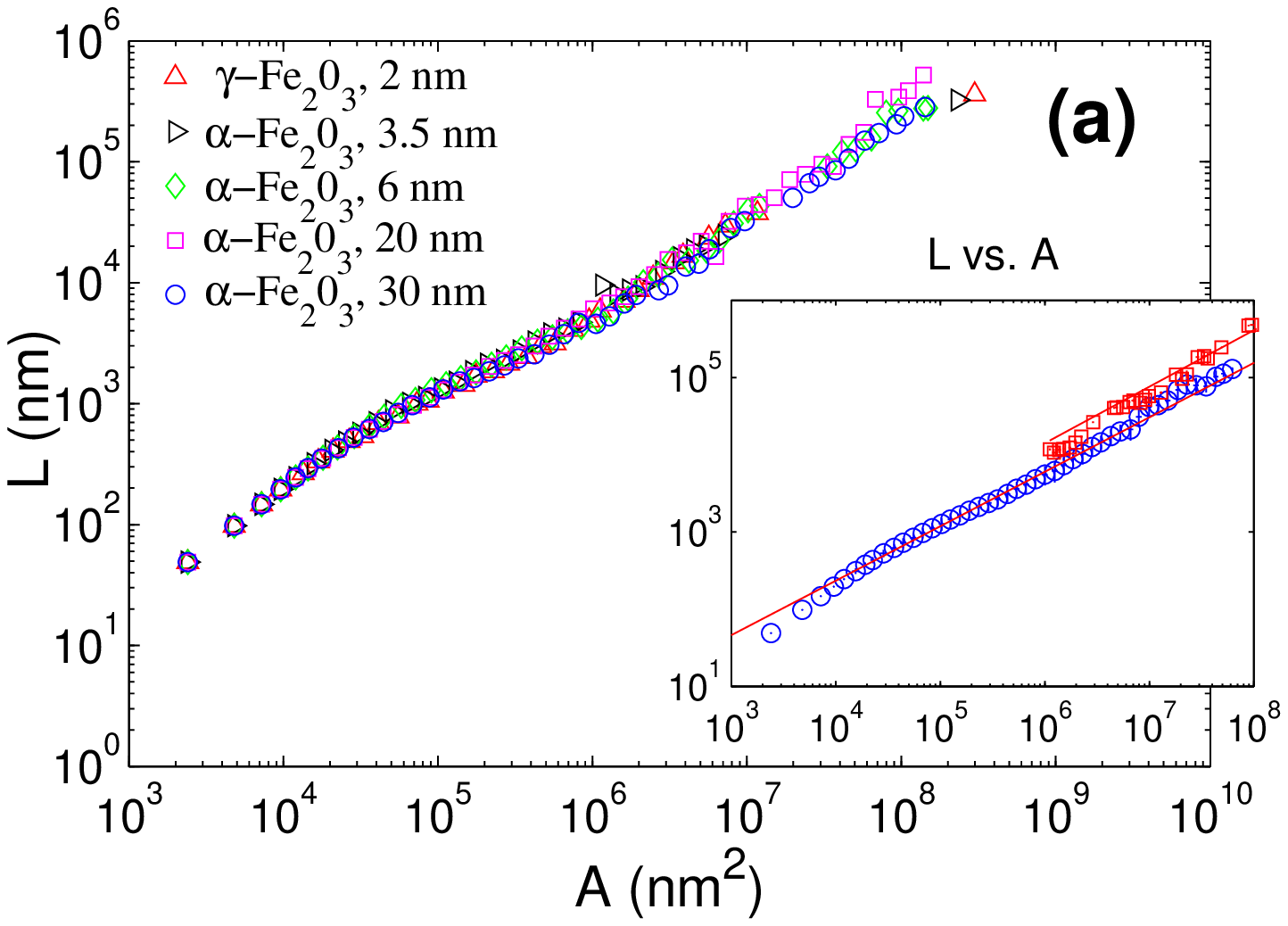}\\
\includegraphics[width=0.75\columnwidth]{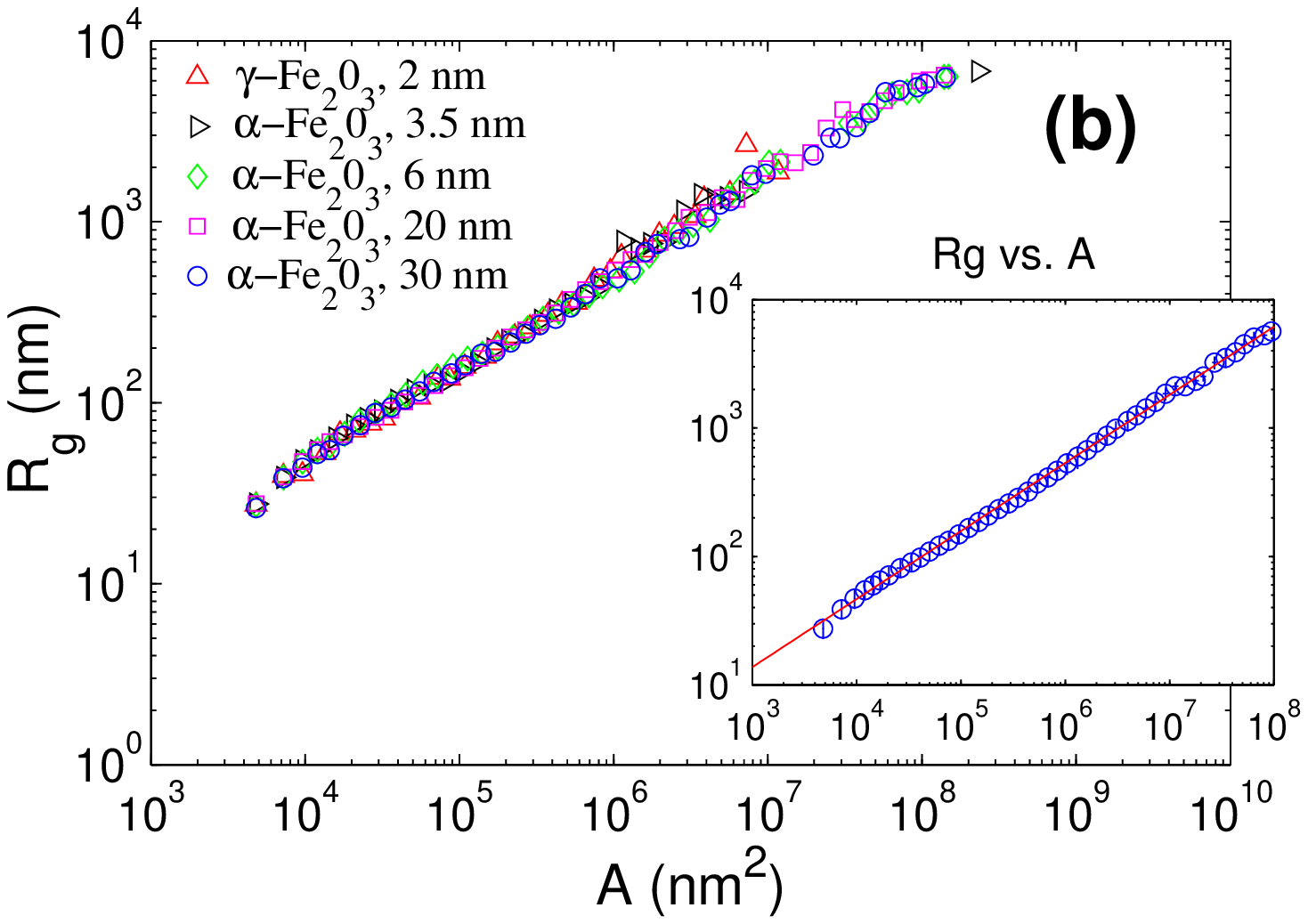}
\caption{(Color online) Morphological features of the magnetic domains of Fe$_2$O$_3$ at various stages of the oxide layer growth: ferrimagnetic $\gamma$-Fe$_2$O$_3$ for thickness $t <$ 3 nm and AF $\alpha$-Fe$_2$O$_3$, $t >$ 3 nm. (a) Domain Perimeters $L$ versus domain area $A$. (b) Domain area $A$ as a function of the domain gyration radius $R_g$. The axes are logarithmic. Insets in both graphs: Average curve over all thicknesses $t$. The power law behavior (straight line) in inset of (a) indicates self-affine domain walls with a roughness exponent $\zeta$=0.60$\pm$0.04. A magnetic field of 2 Tesla was applied to the 2 nm thick $\gamma$-Fe$_2$O$_3$ before the ferri-AF transition. The growth was then pursued up to a thickness of 10 nm. The corresponding data are plotted in red (upper line). The power law behavior (straight line) in inset of (b) indicates a fractal domain dimension $d$$_f$=1.89$\pm$0.02. For both exponents, $\pm$ indicates errorbars for a 95\% confident interval.}
\label{stat1}
\end{figure}

 Figures \ref{general}-a, \ref{general}-b, and \ref{general}-c show typical X-PEEM images of magnetic domains for three important layer thicknesses, 2.5 nm ($\gamma$-Fe$_2$O$_3$), 10 nm and 20 nm ( $\alpha$-Fe$_2$O$_3$), respectively. At first glance, although obtained through two different contrast methods (XMCD and XMLD), these three images seem closely similar. Quantitatively comparing different samples is difficult. Beyond this first observation, we have used a statistical approach to quantitatively characterize the morphology of these magnetic domains.
Perimeter $L$ and gyration radii $R_g$ \cite{noteRg} of domains derived from images of Fe$_2$O$_3$ layers with different thicknesses are plotted as a function of their area $A$ on Figs. \ref{stat1}a and \ref{stat1}b, respectively. In both graphs, the curves obtained for various layer thicknesses below and above the ferri to AF transition are found to overlap almost perfectly without any rescaling factor. In other words, the domains morphology is shown to be statistically invariant with the layer thickness, independently of the magnetic spin ordering, as conjectured from the direct observation of the images in Fig. \ref{general}. This observation is not obvious since for example ferroelectric domains were reported to exhibit size dependence with layer thickness \cite{Catalan:Phys.Rev.Lett.2008}.

To increase the statistics and determine precisely the morphological scaling features of our system, we made use of the statistical invariance with layer thickness and gathered all the data into a single realization set. Domain perimeter $L$ is found to scale as $A^{\sigma}$ with a surface exponent  $\sigma$=0.70$\pm$0.02 \emph{over almost five decades} (Fig. \ref{stat1}b:inset). Since these domains are dense objects, this non-integer exponent is attributed to self-affine scaling features of the boundary. Then, calling $\zeta$ the roughness exponent of the magnetic walls and $\ell$ a unit scale, we expect $L$ to scale \cite{Feder1988} as $\ell^{2-\zeta}$ and A as $\ell^2$. Consequently, $L$ scales as $A^{\sigma=(2-\zeta)/2}$  which leads here to  $\zeta$=0.60$\pm$0.04. Such a roughness exponent is very close to the value $\zeta$=0.633 predicted by the one dimensional (1-D) Kardar Parisi Zhang (KPZ) \cite{Kardar:Phys.Rev.Lett.1986} equation with quenched noise which describes the evolution of an interface in a wide range of systems, e.g. propagation of combustion fronts \cite{Maunuksela:Phys.Rev.Lett.1997}, rupture of paper \cite{Kert'esz:Fractals1993} or flow invasion in porous media \cite{Buldyrev:Phys.Rev.A451992}$\ldots$ Finally, the gyration radius $R$$_g$ was found to scale as $A$$^{1/d_f}$ with the fractal dimension $d_f$=1.89$\pm$0.02 over almost five decades. This value is very close to the fractal dimension $d_f$=91/48 exhibited by connected clusters in the vicinity of the transition in two dimensional (2-D) standard percolation theory \cite{Stauffer1992}.

\begin{figure}
\includegraphics[width=0.75\columnwidth]{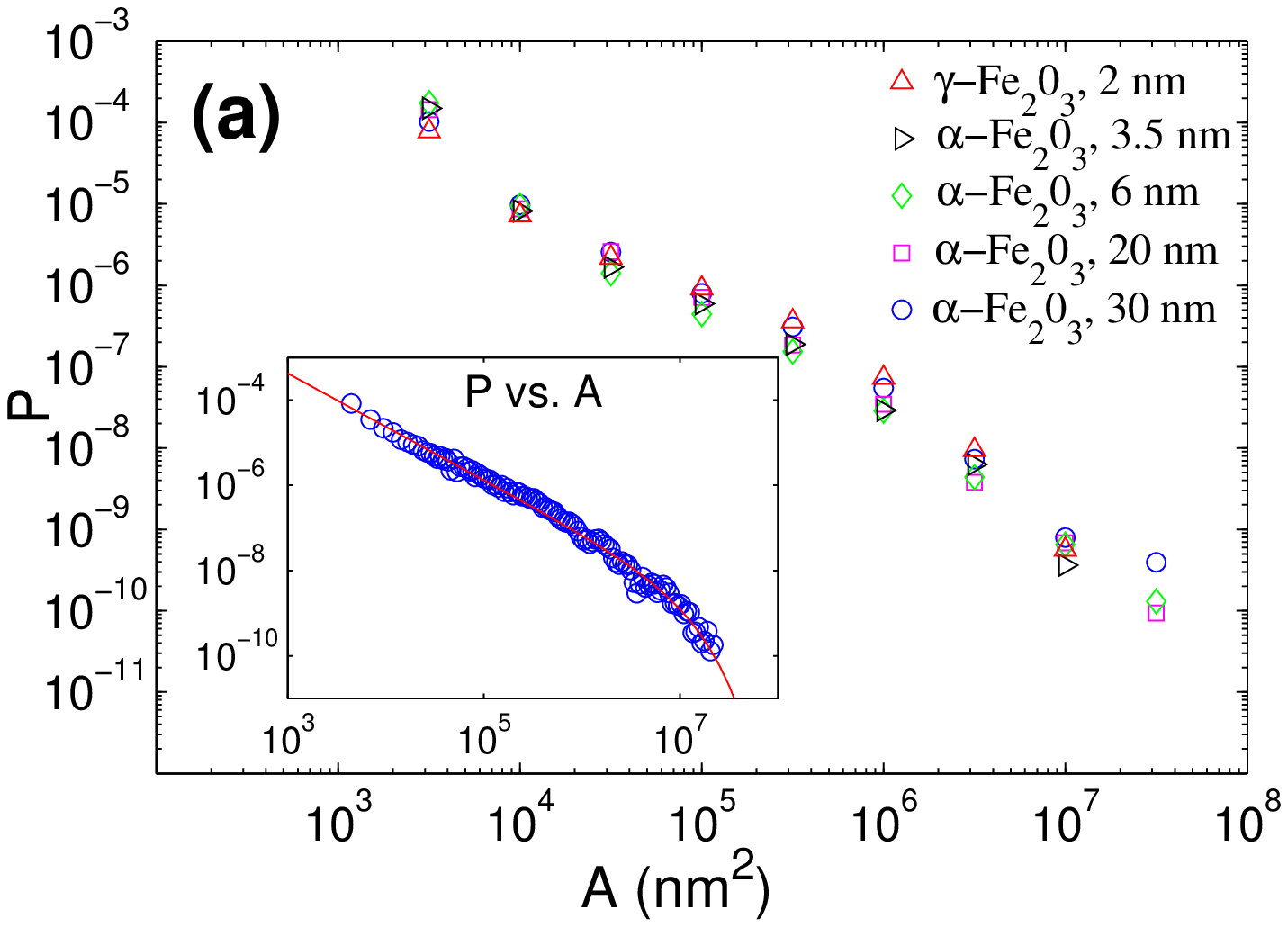}\\
\includegraphics[width=0.75\columnwidth]{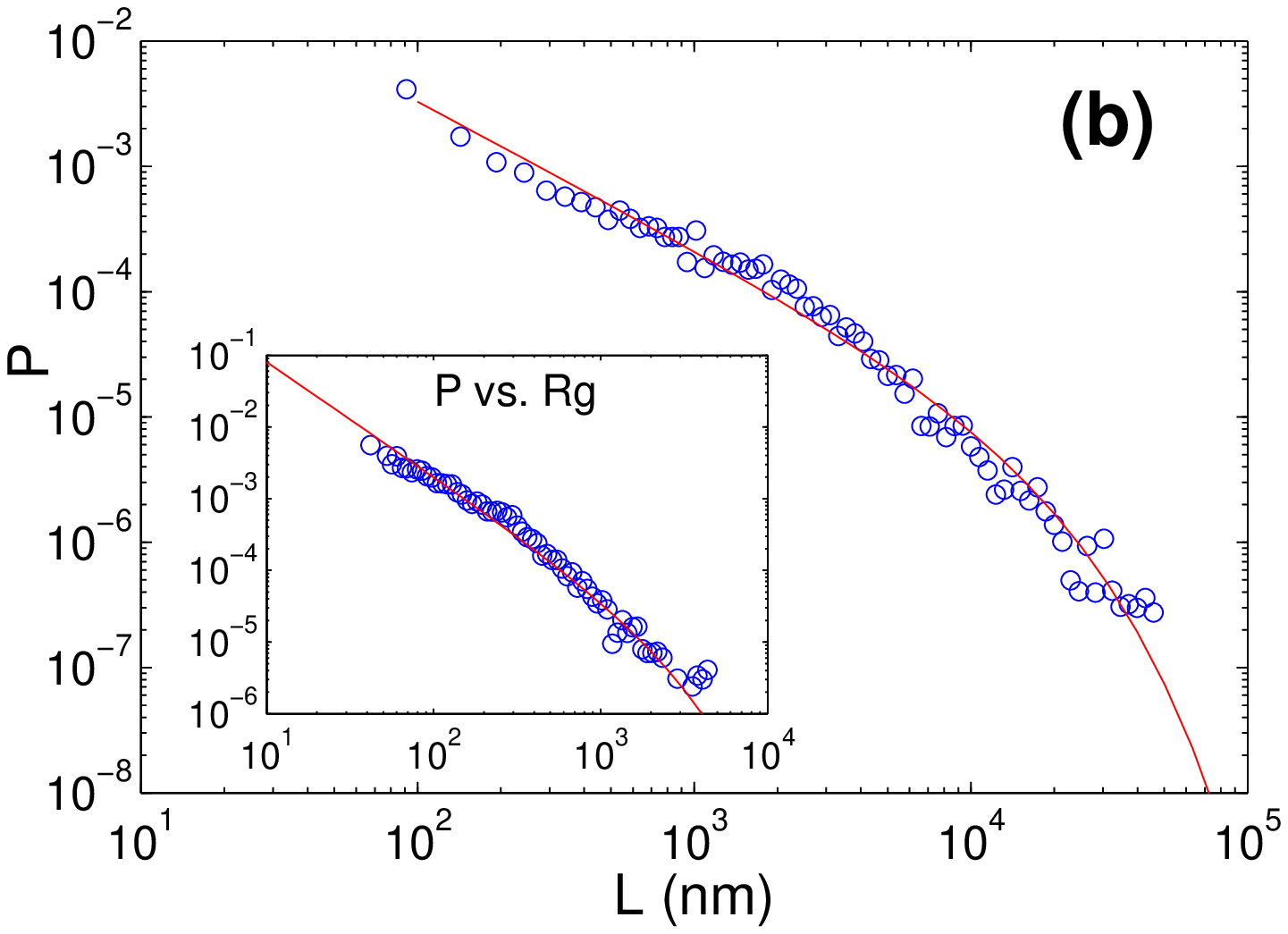}
\caption{(Color online) (a) Main: Probability density function $P$ of the domain area $A$ at various stages of the oxide layer growth: ferrimagnetic $\gamma$-Fe$_2$O$_3$ for thickness $t <$ 3 nm and AF  $\alpha$-Fe$_2$O$_3$ $t >$ 3 nm. Inset: Same curve obtained after having gathered all available data, independently of $t$, so that statistics is increased. (b) Probability density function $P$ of the domain perimeter $L$ (main) and gyration radius $R_g$ (inset) obtained on gathered data. The red lines in (a):main, (b):main and (b):inset correspond to fits $P(A)\varpropto A^{-\tau}\exp(-A/A_0)$, $P(L)\varpropto L^{-\beta}\exp(-L/L_0)$ and $P(R_g)\varpropto R_g^{-\alpha}\exp(-R_g/R_{g0})$, with $\tau=1.25\pm0.04$, $\beta=1.24\pm0.09$ and $\alpha=1.59\pm0.24$, and $A_0=8.1\pm1.1\times10^6\,\mathrm{nm}^2$, $L_0=1.40\pm0.39\times 10^4\,\mathrm{nm}$ and $R_{g0}=2.3\pm1.4\times10^3\,\mathrm{nm}$, where all error bars stand for a 95\% confident interval.}
\label{stat2}
\end{figure}

To complete the statistical analysis of our images, we have computed the distributions of the domains area $A$, perimeter $L$ and gyration radius $R_g$ (Fig. \ref{stat2}). As for the morphological analysis presented in the preceding paragraph, no effect of the layer thickness was evidenced (Fig. \ref{stat2}a:main). This allows gathering all the data to increase the statistics. All these quantities were found to be power-law distributed with an exponential cut-off, the characteristic scale which sets the upper limit of scale invariant morphological features. The exponents $\tau$, $\beta$ and $\alpha$ corresponding to $A$, $L$ and $R_g$ respectively, were found to be: $\tau$=1.25$\pm$0.04, $\beta$=1.24$\pm$0.09 and $\alpha$=1.59$\pm$0.24. It should be emphasized that these exponents are related: since $A$ scales as $R_{g}^{d_f}$, $P(A)$ decreases as $A^{-\tau}$, the distribution $P(R_g)$ goes as $R_{g}^{-(1-d_f(1-\tau))}$  and therefore: $\alpha$=1-$d_f$(1-$\tau$). Similar argument allows to link $\beta$ and $\tau$: $\beta$=($\tau$-$\zeta$/2)/(1-$\zeta$/2). These two relations are perfectly fulfilled indicating a good quality of our statistic data. Hence, the knowledge of a single exponent, - say $\tau$ -, is sufficient to describe the distribution of both the $A$, $L$ and $R_g$.

\begin{figure}
 \includegraphics[angle=0, width=8.5cm]{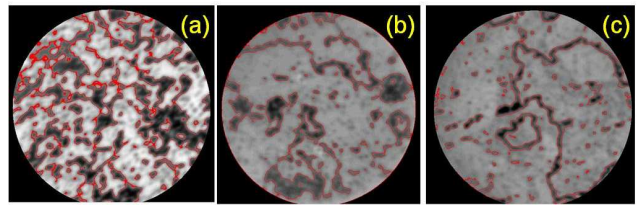}%
\caption{\label{Domains}(Color online) AF domain structure of a 13 nm thick $\alpha$-Fe2O3/Pt(111) film having experienced different magnetic hystories during growth. The field of view is 20 $\mu$m. At 2 nm thickness (a) was demagnetized (without air exposure) using 40 cycles in a magnetic field of $\pm$ 15 mT and (c) was exposed during 10 minutes to a 2 T saturating magnetic field. Sample (b) was grown without magnetic field. Straight lines highlight the domain boundaries. }
\end{figure}

Figures \ref{stat1} and \ref{stat2} fully characterize the distributions and the morphology of the magnetic domains and their boundaries in our system for various thicknesses. The perfect overlap of all the curves demonstrates that the domain structure in the AF phase is (i) statistically invariant with the layer thickness and (ii) identical to the one of the parent ferrimagnetic phase. The first result is highly non-trivial: In FM layers, the mean domain size is known to increase as the square root of the layer thickness \cite{Bataille:Phys.Rev.B2006} and this behavior has been suggested to be more general since recent experiments reveal a similar scaling in AF BiFeO$_3$ films \cite{Catalan:Phys.Rev.Lett.2008}. The second result suggests that the AF and ferrimagnetic domain structures are closely linked. To confirm experimentally this conjecture we applied a saturating magnetic field of 2 Tesla to the 2 nm thick $\gamma$-Fe$_2$O$_3$ layer before the ferri/AF transition. The growth was then pursued with the sample left in a remanent state up to a thickness of 10 nm. From the partial demagnetization process, one expects the smaller domains to disappear. X-PEEM imaging (Fig. \ref{Domains}c) and statistical analysis were carried out. The resulting variation of the perimeter as a function of the domains area is plotted (in red) in fig. \ref{stat1}c. The exponent $\sigma$ remains unchanged, but the prefactor \emph{increases} significantly, shifting the curve in logarithmic scales from the merged curves obtained in all previous situations. This definitively demonstrates experimentally that the AF domain structure of Fe$_2$O$_3$ is inherited from the FM parent layer. The absence of FM parent phases in perovskytes like BiFeO$_3$ films \cite{Catalan:Phys.Rev.Lett.2008} is also consistent with this analysis.

Different magnetic treatments were tested during growth. Figure \ref{Domains} presents the corresponding XMLD X-PEEM images. A 2 nm thick ferrimagnetic  $\gamma$-Fe$_2$O$_3$ layer was grown and exposed to a demagnetization process (Fig. \ref{Domains}a) or to a saturating magnetic field (Fig. \ref{Domains}c) prior the addition of 11 nm of Fe$_2$O$_3$ grown without magnetic field. Compared to an unmodified layer (Fig. \ref{Domains}b) grown without exposure to a magnetic field, demagnetization obviously brakes up the magnetic domains and magnetic saturation promotes larger domains. Importantly, the resulting domain structure is not linked to crystalline parameters. Although magnetic exchange coupling may have a role in the transmission of magnetic domain configuration, the coexistence of FM and AF layers, is here not required. This property may be used to tune the magnetic structure and the domain wall configuration of an AF layer in spintronics devices \cite{bezencenet2010} without requiring any additional patterning process \cite{Takamura:NanoLett.}.

It is now worth discussing the values of the various experimentally measured exponents $\sigma$, $d_f$, $\zeta$ and $\tau$. The morphological scaling features of clusters of aligned spins were computed through Monte Carlo simulations in 2D Ising Models (IM) \cite{Cambier:Phys.Rev.B1986} and 2D Random Field Ising Models (RFIM) \cite{Cambier:Phys.Rev.B1986a} below the critical temperature. The exponents $\sigma$ and $d_f$ were found to be $\sigma$=0.68$\pm$0.04 and $d_f$=1.90$\pm$0.06 in both models, i.e. very close to the ones observed experimentally here. Moreover, it was shown that the propagation of the domains walls in RFIM can be described by the 1-D KPZ equation with quenched noise \cite{Devillard:Europhys.Lett.1992}. In this respect, the values of $\sigma$, $d_f$ and $\zeta$ characterizing the morphology of the magnetic domains structure in Fe$_2$O$_3$ thin films appear natural in the ferrimagnetic phase. On the other hand, it was unexpected that these models would apply to the AF phase observed here for thicknesses larger than 3 nm. This is a direct consequence of the morphological inheritance from the ferrimagnetic parent layer. It appears as a general behavior of thin AF films: applying the same statistical approach to characterize previously published PEEM images of 40 nm thick LaFeO$_3$ films leads to $d_f$=1.90$\pm$0.02 and $\zeta$ =0.58$\pm$0.04 \emph{i.e.} similar exponents \cite{Nolting:Nature2000}. Let us finally mention that the value of $\tau$=1.25$\pm$0.04 obtained here is significantly smaller than the $\tau$=2.05$\pm$0.04 observed in 2D IM \cite{Cambier:Phys.Rev.B1986}, but compatible with observations performed in 2D RFIM where $\tau$ is found to decrease with the strength of the disorder in the magnetic structure \cite{Esser:Phys.Rev.B1997}.

In summary, the statistical characterization of X-PEEM images allowed us to determine the origin of the AF domain structure in $\alpha$-Fe$_2$O$_3$. It is properly dictated by the domain structure of the parent ferrimagnetic $\gamma$-Fe$_2$O$_3$ phase. Any modification of this parent domain structure reflects on the final AF layer, independently of its thickness. This suggests a new and promising route to tailor the magnetic domain wall configuration of AF thin films having ferro- or ferri-magnetic parent phases that are major components of modern spintronic devices.

We are grateful to the SLS-SIM and ELETTRA-Nanospectroscopy beamline (FRB) staffs whose efficient efforts have made these experiments possible. The research leading to these results has received funding from the European Community Seventh Framework Programme (FP7/2007-2013). We thank Luc Barbier for many fruitful discussions and reading the manuscript.


\begin{thebibliography}{35}
\bibitem{noteRg}
The gyration radius is defined here as $R_g = \left\langle (x-\langle x\rangle)^2 +(y-\langle y\rangle)^2  \right\rangle^{1/2}$ where the average $\langle \rangle$ is taken over the positions $(x,y)$ of all pixels belonging to the considered domain.
\end{thebibliography}
\end{document}